# PARALLEL GATED NEURAL NETWORK WITH ATTENTION MECHANISM FOR SPEECH ENHANCEMENT

*Jianqiao Cui, Stefan Bleeck*

Institute of Sound and Vibration Research, University of Southampton, UK

## ABSTRACT

Deep learning algorithm are increasingly used for speech enhancement (SE). In supervised methods, global and local information is required for accurate spectral mapping. A key restriction is often poor capture of key contextual information. To leverage long-term for target speakers and compensate distortions of cleaned speech, this paper adopts a sequence-to-sequence (S2S) mapping structure and proposes a novel monaural speech enhancement system, consisting of a Feature Extraction Block (FEB), a Compensation Enhancement Block (ComEB) and a Mask Block (MB). In the FEB a U-net block is used to extract abstract features using complex-valued spectra with one path to suppress the background noise in the magnitude domain using masking methods and the MB takes magnitude features from the FEB and compensates the lost complex-domain features produced from ComEB to restore the final cleaned speech. Experiments are conducted on the Librispeech dataset and results show that the proposed model obtains better performance than recent models in terms of ESTOI and PESQ scores.

## 1. INTRODUCTION

Single-channel speech enhancement (SE) aims to restore target speech corrupted by background noise. Additive noise degrades the performance of speech recognition systems [1] as well as humans, specifically hearing impaired [2]. Nowadays analytical methods such as Wiener filtering [3] or statistical model-based methods [4] have been replaced with deep neural networks (DNNs) which have already demonstrated promising performance on single-channel speech enhancement [5,6]. Most SE algorithms are either based on mapping [7] or masking [8]. The mapping-based methods mainly use the spectral magnitude or complex-valued features as input [7]. Successful masking-based methods are the ideal binary mask (IBM) [9] or more often the ideal ratio mask (IRM) [10]. For the former, the magnitude and phase information are used individually in the complex domain and estimated to restore the clean speech. For the latter, the original phase information is directly used to reconstruct the output. Often the mean square error (MSE) and scale-invariant SNR (SI-SDR) [11] are adopted as the loss function of the DNNs, however, the speech quality is hard to be estimated as it only weakly correlates with human ratings [12].

Recently, cascaded, or multi-stage concepts have been suggested for SE [13] because the intermediate priors can boost the optimization by decomposing the original task into several sub-tasks. However, each sub-model's performance is restricted because they each only incrementally improve the SNR. In [13], a two-pipeline structure was suggested, using first a coarse spectrum method and secondly a compensating and polishing method. However, the performance of the second part heavily depends on its previous output, and therefore in such a cascade topology, the second-stage model should have enough tolerance to correct for the previous stages' error.

In this paper, we propose a parallel structure for coarse and refined estimation respectively using two modules. The first module (Compensation for Complex Domain Network (CCDN)) calculates masked features to compensate complex components from the second module. In a parallel-path structure, one path is fed with the magnitude spectrum and estimates a mask, the second path outputs complex domain details. Because the mask path deals only with magnitude information, some spectral details will be lost. Li et al. [14] showed that it is important to decouple magnitude and phase optimization. We introduce the compensation path to remove distortion and to compensate lost details. Additionally, in our model we use a module extracting more abstract feature details for the next estimation.

The rest of this paper is organized as follows. Section 2 introduces the signal model. Section 3 introduces our proposed model. In section 4, we present the dataset and experimental setup. The experimental results and comparisons are shown in section 5. Section 6 draws conclusions.

## 2. SIGNAL MODEL FORMULATION

Single-channel speech enhancement aims to remove the background noise $n$ from the single-channel noisy speech $y$, and the corresponding original clean speech denotes $x$.

$$y[t] = x[t] + n[t] \quad (1)$$

Where $t$ represents the time sample index. Meanwhile, we use the short-time Fourier transformation (STFT) to convert the time domain speech signals into time-frequency (TF) domain, that is:

$$Y_{t,f} = X_{t,f} + N_{t,f} \quad (2)$$

Where $y, x, n$ are transformed as $Y, X, N$ by STFT, respectively. $t$ is the corresponding time index and $f$ is the frequency bin. Eq. 2 can also be written as

$$Y_{r\,(t,f)} + jY_{i\,(t,f)} = (X_{r\,(t,f)} + N_{i\,(t,f)}) + j(X_{r\,(t,f)} + N_{i\,(t,f)}) \quad (3)$$

where subscripts $r, i$ respectively represent the real and imaginary part of the complex-valued feature. In the rest content, the $(t, f)$ will be dropped.

## 3. METHODOLOGY

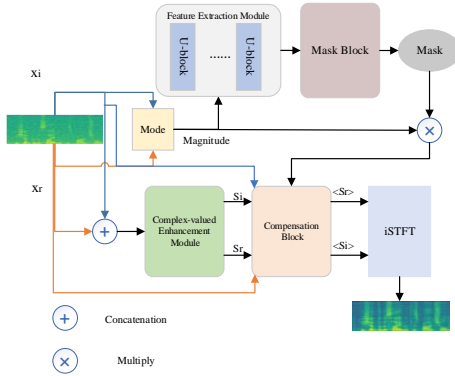

Fig. 1. Architecture of the proposed Compensation for Complex Domain Network

The overall diagram of our proposed model is shown in Fig.1. It consists of 4 parts, Feature Extraction Block (FEB), Mask Block (MB), Complex-valued Enhancement Block (ComEB) and Compensation Block (CB). The model input is the noisy complex spectrum, denoted as $X = Cat(X_r, X_i) \in R^{T*F*2}$, and the corresponding target output is $S = Cat(S_r, XS_i) \in R^{T*F*2}$, where $Cat$ represents the concatenation operation, and $T, F$ denote the time frames as well as frequency bins respectively. Subscripts $r, i$ denote the real and imaginary parts.

### 3.1 Feature Extraction Block

U-nets have been shown to be successful in acoustic feature extraction [15], however, consecutive up and down sampling causes the loss of spectral information. For instance, the power spectral density of harmonic structure from low to high frequency regions will gradually attenuates. For another, the correlation between adjacent frame is important, so it is considerable to obtain both local and global information of each speech sample. U2net [16] was proposed in 2020, whose sub-Unet was employed as embedding layer with residual learning, so as to learning more multi-scale features effectively. Motived by this, in this paper, we replace the traditional 2-D convolutional layer by our proposed U-block module where we use LSTM as the middle layer, to mitigate the information loss, as shown in Fig 2. Fig 3. Shows the details of FEB, it is comprised of Gated Linear Unit (GLU), Layer Normalization (LN), ELU activation function and U-block with residual connection. This structure has 2 advantages, one is the U-block can grasp multi-scale information between frames, which means better abilities to capture contextual features. The other one is that the 2-D GLU can filter some disturbing information and keep useful details. The progress of FEB is given by:

$$y = GLU(x) + U\_block(GLU(x)) \quad (4)$$

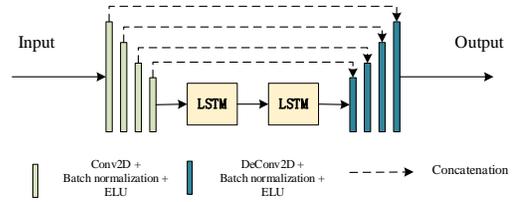

Fig. 2. Architecture of proposed U-block

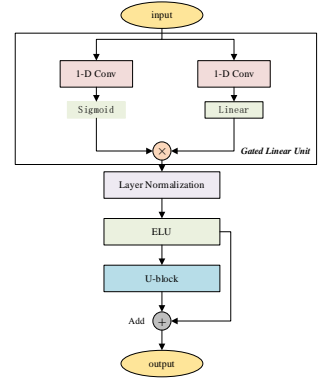

Fig. 3. Architecture of proposed FEB module

### 3.2 Mask Block (MB)

In MB, the output is a mask to suppress the noise in magnitude domain, contributing to coarse features filtered by masks toward to the overall spectrum. Fig. 4 shows the structure, the Mask Block (MB), which consists of encoder, decoder and stacked Gated Residual Units (GRUs), as shown in Fig.3. We use 5 sub-layers in the encoder and decoder respectively. The encoder's sub-layer contains a 1-D convolutional layer with Batch Normalization and utilizes an ELU activation function. In the decoder, the 1-D convolution layer is replaced by a 1-D transposed convolution layer. In the middle layer, a gated residual network is formed by stacked gated residual modules. Novel in this study is that in each GRU we use Multi-head self-Attention (MHSA) in the frequency and time dimension respectively. We test here the hypothesis, based on other work that has been done using MHSA, that this structure can also significantly improve the acoustic receptive field while using only relatively few parameters, while at the same time improving the processing ability of the model to sequential temporal information. In addition, skip connections are used in the gated residual network, which allows the network to incorporate (Add) features extracted from the corresponding layers into the final prediction. Inspired by [17], we implement ISTFT through convolutional layers, so that the time-domain enhanced speech can be used for further training. Moreover, the phase of the original time-domain speech can compensate for the noisy speech phase and reconstruct a more accurate time-domain enhanced speech.

MHSA is often used to extract long term sequence information [18]. MHSA takes as input an L-length sequence feature and produces an output sequence of the same size. This attention mechanism can be described by

$$Attention(Q, K, V) = Softmax\left(\frac{QK^T}{\sqrt{L}}\right)V \quad (5)$$

Where $T$ represents the transpose symbol. The input shape is [Time_frames * Frequency_bins] from encoder, then the proposed MHSA for frequency reshapes the input into [Frequency_bins * Time_frames], as shown in Fig 5(b). Each sub-MHSA can map information along their own specific axis. At the same time, 2 sub-MHSA mechanisms are constructed in parallel and finally their outputs are concatenated together with the original input as the input for the next step. The MHSA for the time frame is described as

$$multi\_head(Q^t, K^t, V^t) concat(h_1, h_2, \ldots, h_t) W_t^{out} \quad (6)$$
Where
$$h_i = Attention(Q^t W_t^Q, K^t W_t^K, V^t W_t^V) \quad (7)$$

Finally, the attention maps are concatenated with the original input and processed by a 1-D convolutional layer to obtain the residual output as the next GRU's input, and is described by:

$$Residual\ output = Conv(MHSA_{time} + MHSA_{freq}) \quad (8)$$

### 3.3 Complex-valued Enhancement Block (ComEB)

The ComEB use almost the same structure as Mask Block (MB), but this block is fed with complex domain features and the GRU does not use attention mechanism replaced by dilation convolution.

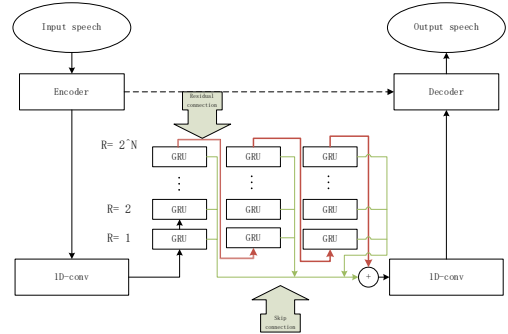

Fig. 4. Architecture of proposed MB

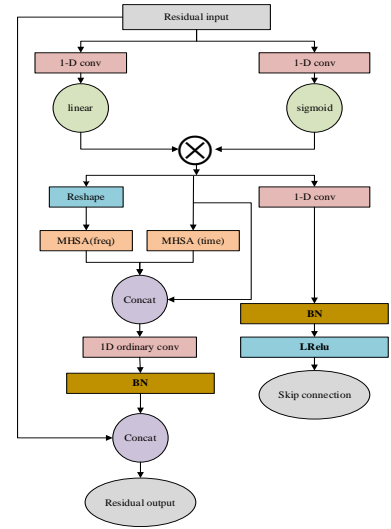

Fig. 5. Architecture of proposed GRU in MB.

Compared with one-dimensional ordinary convolution, one-dimensional dilated convolution can obtain a larger receptive field [19]. The receptive field often grows exponentially, which means that the convolution process can obtain richer speech context feature information and can better mine the information dependencies in the sequence. Meanwhile, the multiple temporal convolutional module (TCM) [20] is also used to leverage the long-range temporal dependencies. Motivated by them, we proposed the ComEB and take both magnitude and phase information into consideration, so as to more effectively alleviate speech distortion.

### 3.4 Compensation Block (CB)

Let $M$ and $X_{com} = \{X_r, X_i\}$ denote the output of MB and ComEB, respectively. Both of them will be fed into CB together with original complex-valued input. As the complement, the RI component as the input and some significant information may be lost by propagation. To update the RI components of the whole model in a collaborative manner, we propose the compensation block acting an important role in this project, shown in Fig 6.

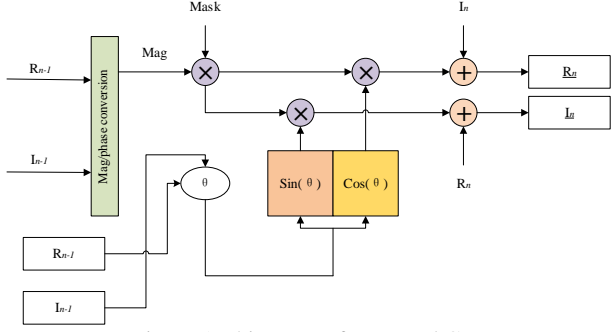

Fig. 6. Architecture of proposed CB

To be specific, the input feature is RI spectrum $\{R_{n-1}, I_{n-1}\}$ firstly decoupled into magnitude spectrum $Mag_{n-1}$, given by:

$$Mag_{n-1} = \sqrt{|R_{n-1}|^2 + |I_{n-1}|^2} \tag{9}$$

$$\theta_{n-1} = arctan2(R_{n-1}, I_{n-1}) \tag{10}$$

Motivated by mask that can effectively and coarsely suppress noise, the $Mag$ will be multiplied by its corresponding mask from MB. However, there is only magnitude feature operated and ignore the phase information, which probably leads to speech mismatch. Besides, values in masks range from 0 to 1 for training stability, both the residual noise and speech distortion will happen accordingly [21]. So, in this case, we design the CB to focus on and compensate the lost detail from the complex domain perspective. The whole procedure is given by:

$$Smag_n = Mag_{n-1} * Mask \tag{11}$$
$$Comp_r = Smag_n * \cos(\theta_{n-1}) \tag{12}$$
$$Comp_i = Smag_n * \sin(\theta_{n-1}) \tag{13}$$
$$R_n = R_{n-1} + Comp_r \tag{14}$$
$$I_n = I_{n-1} + Comp_i \tag{15}$$

Where $*$ means element-wise multiplication operation.

### 3.5 Loss function

The experiments in [22] show that the mean absolute error function (MAE) is improving. The performance is better in terms of speech quality and intelligibility. The frequency domain loss function expression based on MAE is as follows:

$$L_{MAE} = \frac{1}{M}\sum_{n=1}^{M}(\|R_n - R_{n-1}\| + \|I_n - I_{n-1}\|) \tag{16}$$

Where $\|.\|$ denotes the absolute value of vector, $(R_{n-1}, I_{n-1})$ and $(R_n, I_n)$ represent the amplitude spectrum vectors of the original speech and the enhanced speech of the $(n-1)_{th}$ and $n_{th}$ frame, respectively. From Eq 13, 15, apart from RI loss, the magnitude constraint is also considered to improve speech quality.

Where $\|.\|$ denotes the absolute value of vector, $X_n$ and $S_n$ represent the waveform vectors of the original speech and the enhanced speech, respectively. From Eq 13, 15, apart from RI loss, the magnitude constraint is also considered to improve speech quality. According to the literature [32], if the speech evaluation index --- Scale-invariant signal-to-distortion ratio (SI-SDR) is used as the training function of the network, the objective index of enhanced speech is significantly will improve. We can calculate SDR as:

$$SI - SDR = 10\log_{10}\left(\frac{\|\alpha S_n\|^2}{\|\alpha S_n - \widehat{S_n}\|^2}\right) \tag{17}$$

$S_n$ and $\widehat{S_n}$ represent the original magnitude and the enhanced magnitude of the $n_{th}$ frame, respectively. α is the weighting factor of pure speech. Finally, At the same time, we jointly optimize MAE and SI-SDR, and the final network optimization function is (Joint):

$$L_{joint} = L_{MAE} + L_{SI-SDR} \tag{18}$$

### 4. EXPERIMENTS

#### 4.1 Datasets

To evaluate the performance of the proposed model, experiments were conducted on the Librispeech corpus [23]. We selected 6500 clean utterances for the training dataset and 400 for the validation dataset, which are created with SNR levels in [-5dB, -2dB, 0dB, 2dB, 4dB, 5dB, 6dB, 10dB]. For the noise set, around 20000 noises were selected from the DNS-Challenge noise set [24] for training. In this project, we use babble noise and factory1-noise that was mixed with the clean speech. Frame size and frame shift for STFT were set to 512 and 256, respectively.

#### 4.2 Implementation setup

In the FEB, the kernel size, channel and stride of GLU are set to 3, 256 and 1, respectively. Meanwhile, the kernel size and stride of U-block are (1, 3) and (1, 2) in the time and frequency axes with channel set to 256. The number of the (en) decoder layers is decided as 5 and there are totally 2 U-Blocks in this paper. In MB, kernel size, channel, stride in (en) decoder layers are 8, 256, and (1, 3) in the time and frequency axes. In terms of GRU, the left 1D-conv kernel size and the right 1D-conv kernel size are 5 and 1, respectively, and their channel is also 256. In MHSA, the number of heads is set to 2. There are 5 (en) decoder layers and 10 GRUs divided into 2 GRU groups in MB. In ComEB, kernel size, channel, stride in (en) decoder layers are 8, 512, and (1, 2) in the time and frequency axes. The setup of the kernel size, stride and channel in GRU in ComEB is the same as that in MB. The middle part of ComEB is comprised of 4 GRU groups, where there are 5 GRUs. In each GRU group, the dilation rates is [1, 2, 4, 8, 16].

All audio data was sampled at 16kHz and extracted by using frames of 512 with frame shift 256, and the STFT used a Hann window. All models are optimized using the Adam-algorithm [25] with a learning rate of 0.001, decaying by half after each epoch.

## 5. RESULTS AND ANALYSIS

### 5.1 Baseline

Two standard metrics are used to evaluate our model and state-of-the-art (SOTA) competitors: PESQ: Perceptual Evaluation of Speech Quality (from -0.5 to 4.5) [30] and ESTOI: Short-time Objective Intelligibility measure (from 0 to 100(%)) [31]. Both metrics are better if higher.

Our results are shown in table 1 and are compared to 5 relevant state-of-the-art (SOTA) models that were selected for similarity of methodology and recentness: GCRN [26], DCRN [27], PHASEN [28], AECNN [29], and ConvTasNet [20]. In GCRN, DCCRN and PHASEN, complex-domain features are used as input features, as well as magnitude spectrum and phase recovery. The input and output of AECNN and ConvTasNet are raw time-domain waveform.

Table 1
Result of all tested models show PESQ, ESTOI values for SNRs between -3 and 6 dB. 'noisy' is the original noisy speech. 'proposed model' is our model. Higher is better.

| Metrics | PESQ | | | | | ESTOI(%) | | | | |
|---|---|---|---|---|---|---|---|---|---|---|
| Test SNR (dB) | -3 | 0 | 3 | 6 | Avg. | -3 | 0 | 3 | 6 | Avg. |
| Noisy | 1.61 | 1.77 | 1.99 | 2.19 | 1.89 | 31.59 | 40.23 | 49.61 | 58.64 | 45.02 |
| GCRN | 2.32 | 2.62 | 2.87 | 3.08 | 2.72 | 59.57 | 68.83 | 75.72 | 80.78 | 71.23 |
| DCCRN | 2.31 | 2.61 | 2.88 | 3.11 | 2.72 | 59.57 | 68.11 | 75.86 | 81.56 | 71 |
| PHASEN | 2.36 | 2.7 | 2.99 | 3.21 | 2.82 | 61.78 | 71.25 | 78.32 | 83.31 | 73.67 |
| AECNN | 2.32 | 2.64 | 2.91 | 3.11 | 2.74 | 62.13 | 71.57 | 78.25 | 83.03 | 73.74 |
| ConvTasNet | 2.26 | 2.52 | 2.76 | 2.96 | 2.63 | 63.88 | 72.21 | 78.49 | 83.22 | 74.45 |
| Proposed model | 2.34 | 2.56 | 2.99 | 3.28 | 2.79 | 64.43 | 75.24 | 80.63 | 84.64 | 76.29 |

Table 1 shows the results of all experiments. 'Noisy speech' is the original noisy speech, and the table shows all results of baseline models and allows comparison with our proposed modes. Our model outperforms all others in both average PESQ and ESTOI. Note that the average values are simple to compare, but they do not tell the whole story, as enhancement also depends on the SNR. All models work better at higher SNRs as the task is easier, so it is noteworthy that our model works specifically well at the more difficult lower SNRs. The improvements can be explained by comparing against different network. Compared with DCCRN, our model delivers an average of 2.6% improvement in PESQ and 7.5% ESTOI. That is because different from [10], our model adopts ComEB together with MB to reconstruct the target, and therefore the magnitude information produced by MB can be used for compensating the lost details of the complex domain in each training step. Also, unlike traditional monaural speech enhancement, multi-head self-attention in both time and frequency dimensions as adopted in MB also improves the results. Compared with ConvTasNet, on average 6.1% and 2.5% improvements are obtained in PESQ and ESTOI respectively. Although the structure of our MB is similar to that in ConvTasNet, the addition of the GRU leads to further improvement. The attention mechanism used in our experiments delivers better results than traditional dilated convolution in term of the ability to capture global and contextual information for long-range speech sequences.

## 6. CONCLUSIONS

We propose here a collaborative model for speech enhancement in complex domain. Specially, we use a parallel topology in conjunction with the MB. One path in our model estimates the masked magnitude by using an attention mechanism to repair lost details of original input features. The ComEB in the second parallel part combines the outputs from the MB and predicts the clean speech complex features. These two paths are working in tandem to compensate and recover the target on different context time scales.

In the current proposed model, the number of hyper-parameters is more than 100 million, which requires more powerful GPU and time for the training step. So, in the near future, we also plan to speed up our model and decrease the scale of parameters. Another important direction is to utilize parallel multi-stage strategy on different blocks to facilitate the spectrum refinement.